\documentclass[aps, prl, twocolumn, superscriptaddress, longbibliography]{revtex4-1}
\usepackage[T1]{fontenc}
\usepackage[latin9]{inputenc}
\usepackage{amsmath}
\usepackage{amssymb}
\usepackage{hyperref}
\usepackage{xcolor}
\usepackage{graphicx}
\usepackage{bm} 
\usepackage{float}
\usepackage[normalem]{ulem}

\usepackage{braket}

\newcommand{\IQOQI}{\affiliation{Institut f\"ur Quantenoptik und Quanteninformation, \"Osterreichische Akademie der Wissenschaften, 6020 Innsbruck, Austria}}
\newcommand{\UIBKEXP}{\affiliation{Institut f\"ur Experimentalphysik, Universit\"at Innsbruck, 6020 Innsbruck, Austria}}
\newcommand{\UIBKTH}{\affiliation{Institut f\"ur Theoretische Physik, Universit\"at Innsbruck, 6020 Innsbruck, Austria}}
\newcommand{\JILA}{\affiliation{JILA, NIST and Department of Physics, University of Colorado, Boulder, Colorado 80309, USA}}
\newcommand{\CTQM}{\affiliation{Center for Theory of Quantum Matter, University of Colorado, Boulder, Colorado 80309, USA}}

\begin{document}

\newfloat{suppfig}{tbh}{losf}
\floatname{suppfig}{Extended Data Figure }

\title{Quantum-enhanced sensing on an optical transition via emergent collective quantum correlations}

\author{Johannes Franke}
\UIBKEXP
\IQOQI

\author{Sean R. Muleady}
\JILA
\CTQM

\author{Raphael Kaubruegger}
\IQOQI
\UIBKTH

\author{Florian Kranzl}
\UIBKEXP
\IQOQI

\author{Rainer Blatt}
\UIBKEXP
\IQOQI

\author{Ana Maria Rey}
\JILA
\CTQM
\email{arey@jila.colorado.edu}

\author{Manoj K. Joshi}
\IQOQI

\author{Christian F. Roos}
\email{christian.roos@uibk.ac.at}
\UIBKEXP
\IQOQI

\begin{abstract}
The control over quantum states in atomic systems has led to the most precise optical atomic clocks to date~\cite{Bothwell2022, Oelker2019,McGrew2018}. Their sensitivity is currently bounded by the standard quantum limit, a fundamental floor set by quantum mechanics for uncorrelated particles, which can nevertheless be overcome when operated with entangled particles. Yet demonstrating a quantum advantage in real world sensors is extremely challenging and remains to be achieved aside from two remarkable examples, LIGO~\cite{ligo2019,VIRGO2019} and more recently HAYSTAC~\cite{Backes2021}. Here we illustrate a pathway for harnessing scalable entanglement in an optical transition using 1D chains of up to 51 ions with state-dependent interactions that decay as a power-law function of the ion separation. We show our sensor can be made to behave as a one-axis-twisting (OAT) model, an iconic fully connected model known to generate scalable squeezing~\cite{Kitagawa1993} and GHZ-like states~\cite{agarwal_atomic_1997,molmer_multiparticle_1999,song_generation_2019,Comparin2022}. The collective nature of the state manifests itself in the preservation of the total transverse magnetization, the reduced growth of finite momentum spin-wave excitations, the generation of spin squeezing comparable to OAT (a Wineland parameter~\cite{Wineland1992,Wineland1994}
of $-3.9\pm 0.3$~dB for only $N = 12$ ions) and the development of non-Gaussian states in the form of atomic multi-headed cat states in the $Q$-distribution. The simplicity of our protocol enables scalability to large arrays with minimal overhead, opening the door to advances in timekeeping as well as new methods for preserving coherence in quantum simulation and computation. We demonstrate this in a Ramsey-type interferometer, where we reduce the measurement uncertainty by $-3.2 \pm 0.5$~dB below the standard quantum limit for $N = 51$ ions. 
\end{abstract}

\date{\today}

\maketitle

\section{Introduction}

Quantum sensors offer the promise of performing metrological tasks at a level not possible in classical systems by harnessing entanglement~\cite{Giovannetti2011, Degen2017, Pezze2018}. Nonetheless, to generate entanglement, interactions are required, which add undesirable complications, particularly when they are short-ranged in nature~\cite{Ludlow2015}. From this consideration, sensors offering full connectivity and operating with macroscopic ensembles to enhance the rate and level of achievable entanglement, such as atoms interacting via all-to-all photon-mediated interactions in optical cavities, are currently at the frontier of entanglement generation~\cite{norcia_cavity-mediated_2018,ritsch_cold_2013,leroux_implementation_2010,hosten_measurement_2016,cox2016deterministic,pedrozo2020entanglement}. However, these platforms lack the desired single-particle control in a quantum sensor, given that size and control of a system are often competing priorities. On the contrary, in platforms where individual addressing is currently possible, including ultracold molecules~\cite{Bohn2017}, optical lattice or tweezer clocks~\cite{Ludlow2015,Schine2022}, 
interactions between atoms typically decrease with distance, and the ability to engineer all-to-all connectivity is often severely limited. In trapped ions, both single-particle control and all-to-all interactions are available. However, implementing all-to-all interactions in arrays of tens to hundreds of ions~\cite{Britton2012, Bohnet2016} faces technical hurdles that limit the coherent generation of entanglement.

\begin{figure*}[ht]
 \centering
 \includegraphics{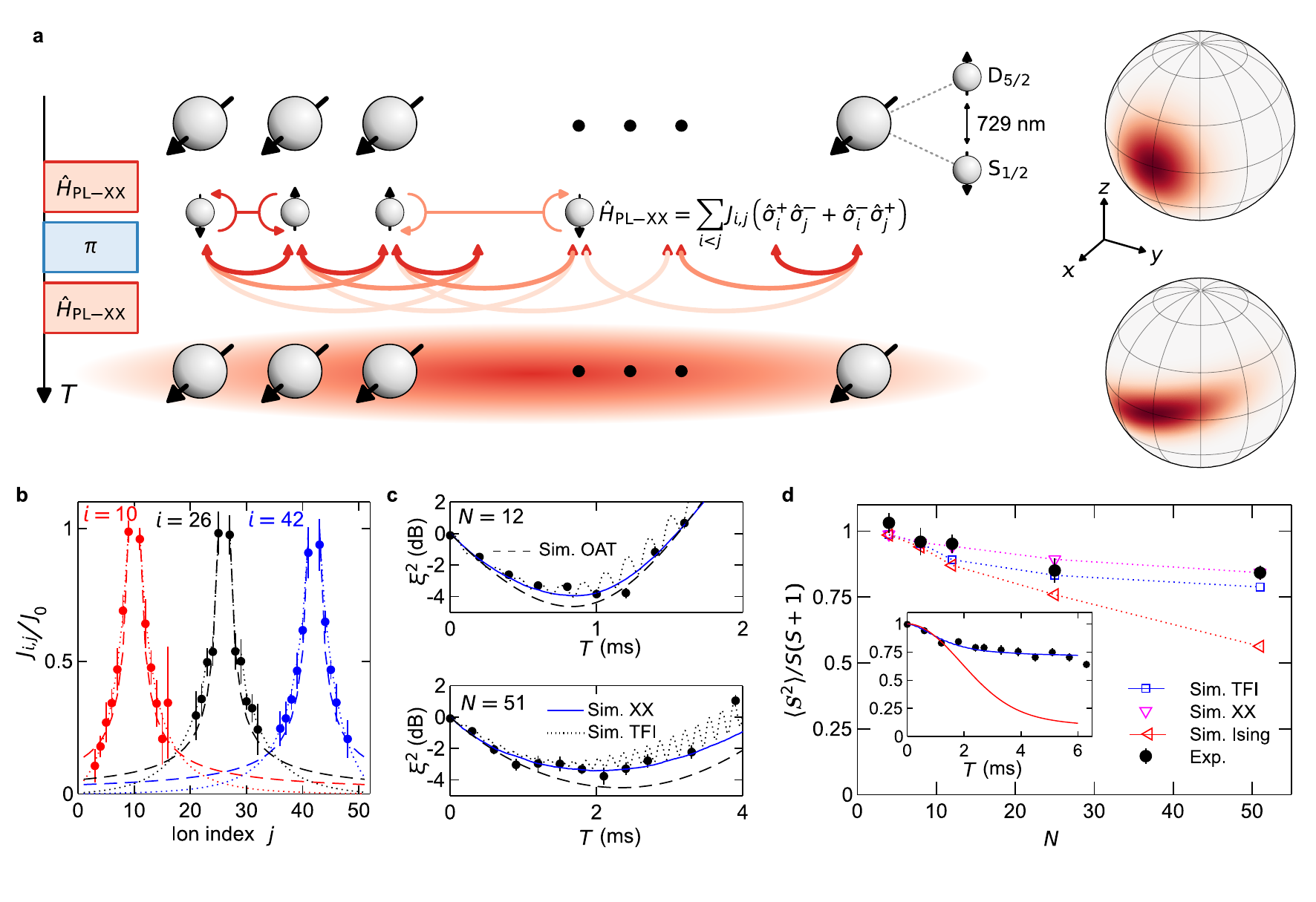}
 \caption{\textbf{Realization of squeezing by non-collective interactions.} a) Experimental sequence describing exchange interaction-induced squeezing of the overall spin vector. We provide a schematic of our ion chain, with qubits encoded on an optical transition. These interact via power-law XX interactions, where the interaction rate decreases with distance. Our sequence combines interaction pulses (red) with pulses from a global addressing beam (blue) to evolve an initial CSS into a SSS, as illustrated on the collective Bloch sphere. b) Measured spin-spin couplings $J_{i,j}$ in a 51-ion chain (dotted line: fit by a power-law interaction with $\alpha\approx0.9$, dashed line: predicted coupling for the given mode structure and laser detuning). c) Measured squeezing vs. interaction time for $\alpha \approx 1$, $J_0=560$~rad/s in a 12-ion chain and $\alpha\approx 0.9$, $J_0=216$~rad/s in an $N=51$ ion chain. We compare to numerical calculations of the dynamics for the corresponding XX model (blue, solid) and dynamics in the presence of a finite transverse field (black, dotted, see methods \ref{met:numerics}), as well as theoretical results for the OAT model (black dashed lines), all taking into account our measured decoherence. d) Total spin, measured at the time at which the spin squeezing peaks in the XX model (interaction strength between 216 and 234~rad/s for all system sizes $N$), and normalized by its maximal value $S(S+1)$ for $S = N/2$. We compare to numerical simulations with (blue, squares) and without (magenta, downward-pointing triangles) a finite transverse field and to analytic results for the Ising model after the same evolution time (red, left-pointing triangles). The inset shows the total spin as function of interaction time $T$ for 51 ions, compared to theoretical results for the TFI (blue) and Ising (red) models.}
 \label{fig:1}
\end{figure*}

A possible solution to harnessing metrologically useful entanglement has been proposed with the suggestion that systems exhibiting short-ranged interactions could nonetheless offer an opportunity for generating entanglement for metrological applications at a level comparable to all-to-all interacting systems~\cite{Perlin2020,bilitewski_dynamical_2021,Comparin2022,young_enhancing_2023,Block2023}, though such predictions inherently rely on uncontrolled approximations to the quantum many-body dynamics since exact treatments are not currently available. Here, we experimentally validate these predictions by transforming the fragile Ising interactions arising between an optical transition in a string of up to 51 trapped $^{40}$Ca$^+$ ions coupled via power-law interactions into an XX model by applying a large transverse drive to our system. In contrast to the Ising model, the XX model features interactions that favor spin alignment, and thus can stabilize collective behavior. We indeed observe that this model enables the survival of features intrinsic to OAT dynamics, including spin squeezing, the development of collective excitations, and, at later times in the dynamics, the creation of multi-headed atomic Schr{\"o}dinger cat states~\cite{agarwal_atomic_1997,molmer_multiparticle_1999,song_generation_2019,comparin_multipartite_2022}. Our work demonstrates that the generation of collective entanglement for metrology can be realized in an array of platforms utilizing finite-ranged interactions validating prior theoretical predictions. Our observations have important implications for a range of emerging quantum simulators and sensors where all-to-all connectivity is not feasible. Furthermore, we show that the relative simplicity of our protocol, compared to more elaborate, and carefully calibrated schemes~\cite{Marciniak2022}, allows its implementation in larger arrays that can achieve improved phase estimation sensitivity in a Ramsey sequence. 

\section{Setup and Theory}
Quantum limits in sensing are commonly discussed for collective systems where all operations act identically on each particle. A paradigmatic interaction that is collective and can be used to entangle $N$ sensor atoms is the OAT model, described by the Hamiltonian 
\begin{align} 
\hat{H}_{\rm{OAT}} = \frac{\chi}{2} \sum_{i < j}^N \hat{\sigma}_i^z\hat{\sigma}_j^z,\label{eq:OAT}
\end{align}
where we define Pauli operators $\hat{\sigma}^x = \hat{\sigma}^+ + h.c.$, $\hat{\sigma}^y = -i(\hat{\sigma}^+ - h.c.)$, with $\hat{\sigma}^+ = \ket{\uparrow}\bra{\downarrow}$, and $\hat{\sigma}^z = \ket{\uparrow}\bra{\uparrow} - \ket{\downarrow}\bra{\downarrow}$. 
The corresponding collective spin operators are $\hat{S}_\mu = \sum_i \hat{\sigma}_i^\mu/2$, $\mu = x,y,z$. This model conserves both the total $z$-magnetization $\hat{S}_z$ as well as the total spin $\hat{\bm{S}}^2$.

Starting from an initial uncorrelated coherent spin state (CSS) with all spins polarized along $+x$ $\ket{\bf{x}} = \otimes_{i=1}^N \ket{+}_i$, and $\ket{+} = (\ket{\uparrow} + \ket{\downarrow})/\sqrt{2}$, the OAT Hamiltonian shears the classical noise distribution and transforms it into a spin-squeezed state (SSS).
The presence of spin squeezing in the resulting state can be quantified through the Wineland parameter, defined as~\cite{Wineland1992,Wineland1994}
\begin{align}\label{eq:Sqparam}
 \xi^2 = \min_{\mathbf{n}_{\perp}} \frac{N\langle (\Delta\hat{S}_{\mathbf{n}_{\perp}})^2\rangle}{|\langle \hat{\mathbf{S}}\rangle|^2},
\end{align}
where the spin variance $\braket{(\Delta\hat{S}_{\mathbf{n}_{\perp}})^2} \equiv \braket{(\hat{S}_{\mathbf{n}_{\perp}} - \langle \hat{S}_{\mathbf{n}_{\perp}}\rangle)^2}$ is minimized over all axes $\mathbf{n}_{\perp}$ transverse to the Bloch vector $\langle \hat{\mathbf{S}}\rangle$. This parameter witnesses entanglement when $\xi^2<1$, and quantifies the metrological gain in phase sensitivity, $\Delta\phi$, to a collective rotation over that achieved by an initial uncorrelated state, $\Delta\phi_{\rm SQL}, =1 / \sqrt{N}$, i.e. $\xi^2 = (\Delta\phi/\Delta\phi_{\rm SQL})^2$.
The OAT dynamics can prepare spin squeezing that ideally scales as $\xi^2\sim N^{-2/3}$~\cite{Kitagawa1993}. At longer evolution times, this model can also generate various non-Gaussian states, such as $q$-headed cat states, whose metrological utility must be characterized through more complex, nonlinear quantities beyond the squeezing parameter~\cite{gessner_metrological_2019}. While the emergence of such states are often considered key features of this collective model, it has been suggested that models lacking this collective symmetry may nevertheless exhibit similar dynamical features of the OAT model, particularly in regards to its metrological utility.

One such model is the power-law Ising chain, accessible in a plethora of modern platforms~\cite{Browaeys2020,Bruzewicz2019,Tscherbul2022,gorshkov_tunable_2011}:
\begin{align}
 \hat{H}_{\rm{PL-Ising}} = \frac{1}{2}\sum_{i<j} J_{i,j}\hat{\sigma}_i^z\hat{\sigma}_j^z,\label{eq:Ising}
\end{align}
where $J_{i,j} = J_0|i-j|^{-\alpha}$ for exponent $\alpha$, resembling $\hat{H}_{\rm{OAT}}$ with the addition of a distance-dependent interaction. Although this Hamiltonian induces a similar shearing effect as the OAT model, it does not preserve the total spin; as a result, the dynamics only approximate those of a collective OAT model, with $\chi = \overline{J}$, for sufficiently long-ranged interactions, where $\overline{J} = \sum_{i<j} J_{i,j}/(N(N-1)/2)$ is the average coupling between pairs of spins. For $\alpha \geq D$,
this model ceases to exhibit scalable spin squeezing among other behaviors characteristic of fully collective systems, and only recovers the full $N^{-2/3}$ scaling when $\alpha < 2D/3$, with $D$ the dimensionality of the system~\cite{FossFeig2016}. This already precludes scalable spin squeezing generation in the case of $\alpha = 1$ relevant for our $D=1$ experiment.

Nonetheless, collective behavior in the power-law Ising model may potentially be stabilized through the appropriate addition of Heisenberg-type interactions $\hat{H}_{\rm{PL-Heisenberg}} = \sum_{i<j} J_{i,j}\hat{\bm{\sigma}}_i \cdot \hat{\bm{\sigma}}_j/2$~\cite{rey_many-body_2008,Perlin2020,Comparin2022,Block2023,young_enhancing_2023}, which conserve total spin for any interaction range. A key example of this is the power-law XX model $\hat{H}_{\rm{PL-XX}} = \hat{H}_{\rm{PL-Heisenberg}} - \hat{H}_{\rm{PL-Ising}}$, which results in distance-dependent exchange interactions of the form $\hat{\bm{\sigma}}_i \cdot \hat{\bm{\sigma}}_j - \hat{\sigma}_i^z \hat{\sigma}_j^z = 2(\hat{\sigma}_i^+ \hat{\sigma}_j^- + h.c.)$. Unlike the integrable OAT and Ising models, where the dynamics of arbitrary correlators can be exactly solved for virtually any $N$~\cite{foss-feig_nonequilibrium_2013}, computing the dynamics of the power-law XX model is generically challenging beyond a couple dozen spins, even in the absence of decoherence. While approximate methods have been developed to tackle the quantum dynamics of such systems~\cite{Schachenmayer2015,Comparin2022} these fundamentally rely on uncontrolled approximations or ansatzes, and thus remain to be validated in an experimental setting.

In our trapped-ion quantum simulator of up to 51 ions, a pseudo-spin is encoded in two electronic states of $^{40}$Ca$^+$ --- the $\ket{\downarrow} = \ket{\mbox{S}_{1/2},m=+1/2}$ and $\ket{\uparrow} = \ket{\mbox{D}_{5/2},m=+5/2}$ states --- which are collectively coupled by a global laser beam~\cite{Kranzl2022b}. Spin-spin interactions between the ions are engineered via a two-tone laser that couples the internal electronic states of the ions to their ground-state cooled transverse motional modes (see methods \ref{met:exp}, \ref{met:interactions}). With the application of a strong drive transverse to the interaction axis, the dynamics are described by the power-law transverse field Ising model (TFI), $\hat{H}_{\rm{PL-TFI}}= \sum_{i<j}J_{ij}\hat{\sigma}_i^x\hat{\sigma}_j^x+B\sum_i \hat{\sigma}_i^z$. By considering this system in the rotating frame of the drive, we can approximately describe it via the power-law XX model
\begin{align}\label{eq:XYInteraction}
\hat{H}_{\rm{PL-XX}} = \sum_{i<j} J_{i,j} \left(\hat{\sigma}_i^+ \hat{\sigma}_j^- + \hat{\sigma}_i^- \hat{\sigma}_j^+\right)
\end{align}
as previously described, where $J_{i,j} = J_0|i-j|^{-\alpha}$ (see Fig.~\ref{fig:1}b) sets the strength of interactions between sites $i$ and $j$. The interactions strength is parametrized in terms of the nearest neighbour strength $J_0$ and a tunable exponent $0 < \alpha < 3$ describing the interaction range.

\section{Results}
\emph{Generation of spin-squeezed states --}
We use a similar protocol for preparing spin-squeezed states as for the OAT, as sketched in Fig.~\ref{fig:1}a: we first prepare a CSS polarized along $+x$ and then evolve this state under the XX interaction in Eq.~\eqref{eq:XYInteraction} for a variable time $T$. In the experiment, the interaction period is split by an echo pulse that cancels site-dependent Stark shifts along $z$ and increases the system's coherence time.

In Fig.~\ref{fig:1}c, we investigate the dynamics of the Wineland squeezing parameter for two system sizes and interaction ranges. We find an optimal value of $\xi^2 = -3.9~\pm~0.3$~dB with $N=12$ and $\alpha \approx 1$, and $\xi^2 = -3.7~\pm~0.5$~dB with $N=51$ and $\alpha \approx 0.9$, comparable to levels of noise reduction generated in prior trapped-ion studies using larger arrays~\cite{Bohnet2016}. Despite the increased particle number in the latter case, we do observe a slight decrease in the attainable spin squeezing for this system, in contrast to the expected improvement according to the ideal OAT model. We attribute this to collective dephasing noise from magnetic field fluctuations and laser noise, at a rate we characterize independently and can be included in our theoretical calculations (see methods \ref{met:echo}, \ref{met:numerics}). To verify that this is the mechanism responsible for the reduction in spin squeezing, we compare the dynamics to that of the OAT model with coupling $\chi = \overline{J}$, also in the presence of this collective dephasing. For both system sizes, we observe a similar reduction in the spin squeezing, and we find that the observed dynamics are within less than $1$ (dB) of the corresponding OAT dynamics, confirming our dynamics are well approximated by the fully collective model, and that indeed any reduction in the expected metrological gain is not a result of the reduced interaction range of our power law model or other local decoherence mechanisms in the experiment.
Furthermore, we compare to analogous calculations of Eq.~\eqref{eq:XYInteraction} with decoherence, finding excellent agreement with the observed dynamics utilizing the assessed decoherence rate. Lingering deviations between the two are small, and largely accounted for by considering the underlying model in the presence of the transverse field, validating both our approximate use of the XX model as well as the validity of the approximate numerical methods we employ (see methods \ref{met:numerics}. Note that an exact calculation of the dynamics of Eq.~\eqref{eq:XYInteraction} or the TFI model, in the presence of decoherence, is currently intractable for the range of $N$ we consider. 

To further probe the collective nature of our system, we examine the value of the total spin $\langle \hat{\mathbf{S}}^2\rangle$ at the time at which the spin squeezing is found to be optimal, for a range of system sizes; see Fig.~\ref{fig:1}d. This observable commutes with the global dephasing operator, and is thus unaffected by it. We observe that total spin decays slightly for larger $N$, as the finite range of the interactions is better resolved over larger chains, but appears to plateau to a relatively large value; this is consistent with numerical calculations of both the TFI and XX models, demonstrating the latter to be well-approximated by our system. Furthermore, when compared to the state prepared by the Ising model with the same interaction range (Eq.~\eqref{eq:Ising}), the observed dynamics in our XX model remain relatively collective, consistent with theoretical predictions~\cite{Perlin2020,Comparin2022,Block2023}.

\begin{figure}[ht]
 \centering
 \includegraphics{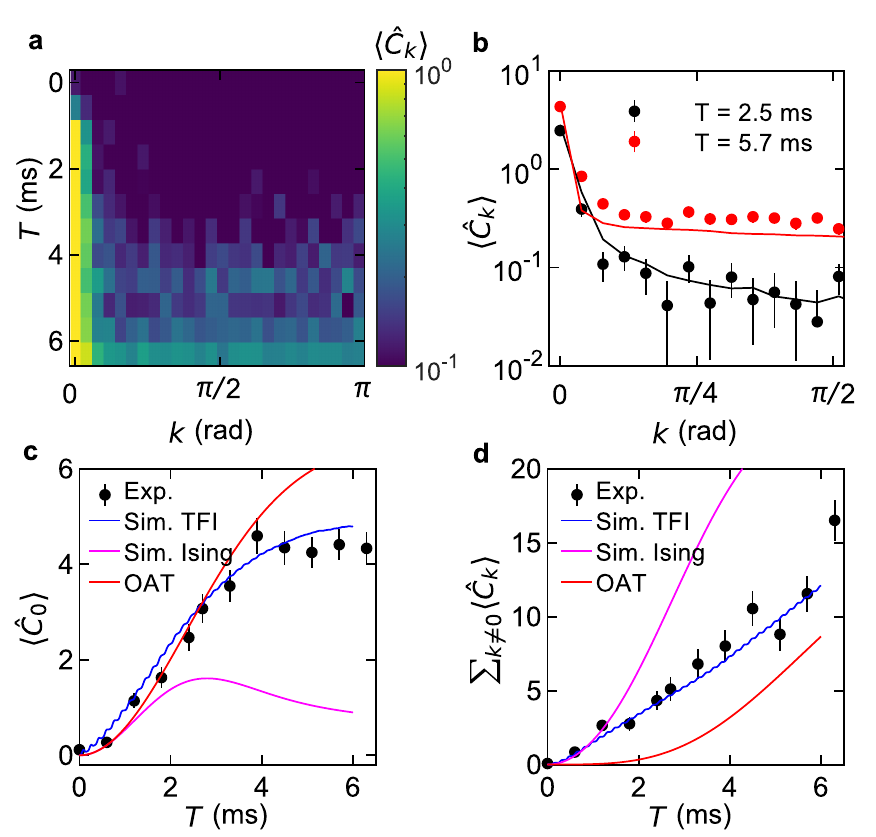}
 \caption{\textbf{Spin-wave propagation.} a) Time evolution of the occupations $\langle \hat{C}_k\rangle$ of linear spin waves with wavevector $k$ in a $N = 51$ ion chain. b) Measured occupations $\langle \hat{C}_k\rangle$ at $T = 2.5$~ms and $5.7$~ms. Solid lines are numerical simulations for the transverse field Ising model (TFI). c) Time evolution of the $k=0$ correlation in comparison to simulated TFI, Ising, and OAT models. d) Summed occupations of the non-collective $k\neq 0$ spin waves. }
 \label{fig:2}
\end{figure}

\emph{Spin correlations --} 
Another way to understand the relatively large spin squeezing we observe in our measurements is to study the spatial structure of spin-spin correlations in the ion chain. Generally, systems exhibiting non-uniform interactions will develop non-collective spin-wave excitations (SWE) characterized by a finite wavenumber (quasi-momentum $\hbar k$) $k\neq 0$. In particular, the spin correlations can be used to estimate the mode occupation $\langle \hat{n}_k\rangle$ of SWE (see methods \ref{met:LSW}). Using the well known 
Holstein-Primakoff approximation~\cite{kurucz2010multilevel}, which maps spin operators into quadratures written in terms of bosonic annihilation and creation operators, the SWE can be connected to two-point spin correlators:

\begin{align}
& \frac{\langle\hat{n}_{k} + \hat{n}_{-k}\rangle}{2} \simeq \braket{\hat{C}_k} \equiv  \frac{1}{2} - \frac{1}{2N} \sum_i \braket{ \hat{\sigma}_i^x} \notag\\
 &\quad + \frac{1}{2N} \sum_{i<j} \braket{\hat{\sigma}_i^y\hat{\sigma}_j^y + \hat{\sigma}_i^z\hat{\sigma}_j^z }\cos(k (r_i - r_j))
\end{align} 
Utilizing site-resolved measurements of our system, we study the dynamics of $\braket{\hat{C}_k}$ under evolution described by Eq.~\eqref{eq:XYInteraction}. In Fig.~\ref{fig:2}a, we measure $\braket{\hat{C}_k}$ as a function of $k$ for various times. We observe a significant growth of the $k=0$ mode, while the populations of other modes with $k\neq 0$ attain a relatively small growth between the two selected times, of which the most significant occupation occurs for small $|k|$ around $0$; see Fig.~\ref{fig:2}b for experimental data and corresponding numerics for the TFI model at selected times.

\begin{figure*}[ht]
 \centering
 \includegraphics{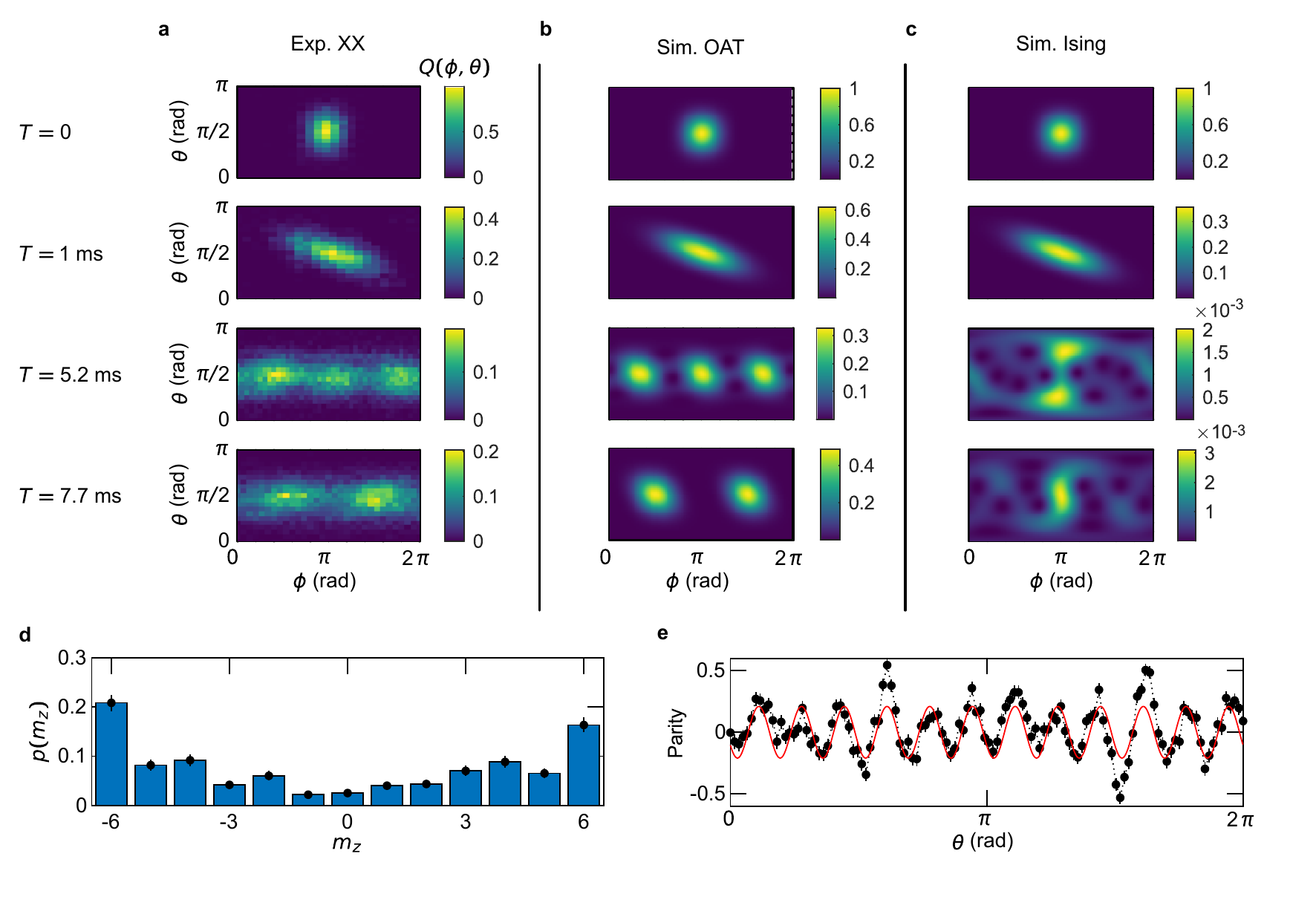}
 \caption{\textbf{Husimi Q-distributions.} a) Measured Husimi Q-distributions of 12-ion spin states for different interaction times $T$ of the realized XX interaction with $J_0=560$~rad/s. While the state evolves from a CSS to a SSS at short times, a 3-headed cat state and a 2-headed cat state are observed at later times. b), c) Simulated Husimi Q-distributions for an OAT model with $\chi = \overline{J}$ and the power-law Ising model with the same interaction range as the experiment, at different interaction times $T$ and without the effects of decoherence. For illustrative reasons, the peaks of the distributions are centered by shifting the phase $\phi$. We adjust the color scale of each plot to enhance the visibility of features of the phase space distributions. d) Measured probabilities in all magnetization sectors $m_z = (n_{\uparrow} - n_{\downarrow}) / 2$ of the 2-headed cat state. Here $n_{\uparrow (\downarrow )}$ is the number of ions in the state $\ket{\uparrow}(\ket{\downarrow})$. e) Parity oscillations of 2-headed cat state and corresponding sinusoidal fit in red to estimate the contrast $C$. The black dotted line is a guide to the eye.}
 \label{fig:3}
\end{figure*}

In Fig.~\ref{fig:2}c and d, we plot the dynamics of both the $k=0$ mode and the total $k\neq 0$ mode populations. While the TFI data features a comparable growth of the collective mode compared to the total occupation of the non-collective modes, in the corresponding power-law Ising model with the same interaction range, the population of the non-collective modes quickly outpaces the growth of the collective mode, signifying a regime where the dynamics is dominated by finite-momenta excitations and where the linear spin-wave approximation breaks down. Furthermore, we find good agreement between the data and corresponding theoretical results for the OAT model, further validating the collective nature of our system. We note that, in contrast to the spin squeezing, the SWE are relatively robust to global dephasing, and we thus directly compare the observed experimental results to theory calculations obtained by solving the ideal unitary dynamics. The slow, finite growth of the $k\neq 0$ population in the OAT model is an artifact arising from the onset of non-Gaussian correlations at times beyond $\sim 2$~ms, which are neglected in the linear SWE, and which we now proceed to examine.

\emph{Q-functions: beyond the Gaussian regime --}
For evolution times beyond the optimal spin squeezing time, the quantum noise distribution in the OAT model starts to develop various non-Gaussian states, a set of which are the family of atomic $q$-headed cat states~\cite{agarwal_atomic_1997,molmer_multiparticle_1999,song_generation_2019,Comparin2022}. These correspond to a superposition of $q$ coherent spin states, with polarizations equally spaced about the equator of the Bloch sphere. Of these states, the multi-headed cat state with $q=2$ state corresponds to the well-known Greenberger-Horne-Zeilinger (GHZ) state~\cite{greenberger_GHZ_1989}. This is a key resource in metrology and error correction protocols, exhibiting maximal depth of entanglement, and can be used in principle to enhance the phase sensitivity of a measurement by a factor of $N$.

A drawback of GHZ states is that they are extremely susceptible to decoherence and inhomogeneities induced by coherent non-collective interactions. While we have demonstrated that XX interactions can help stabilize collective dynamical behaviors for time scales required to prepare Gaussian spin-squeezed states, far longer time scales are required to prepare GHZ states. 

To explore and characterize the survival of the collective non-Gaussian states in the dynamics under Eq.~\eqref{eq:XYInteraction}, we directly measure the evolution of the Husimi Q-distribution  $Q(\theta,\phi) = \langle \bf{n}(\theta,\phi)|\hat{\rho}|\bf{n}(\theta,\phi)\rangle$, where $\hat{\rho}$ is the density matrix of our state and $|\bf{n}(\theta,\phi)\rangle$ is the CSS with polarization vector $\bf{n}$ characterized by polar/azimuthal angles $\theta/\phi$ on the collective Bloch sphere. In Fig.~\ref{fig:3}, we show the measured Q-distribution for various times in the evolution. Starting from the initial CSS at $T=0$, we initially observe the development of a SSS at short times. At specific later times, we observe the fracturing of the Q-distribution into various distinct patches, with the distribution characterized by $q$ such patches occurring at time $T = \pi/q\overline{J}$.

\begin{figure}[!htb]
 \centering
 \includegraphics{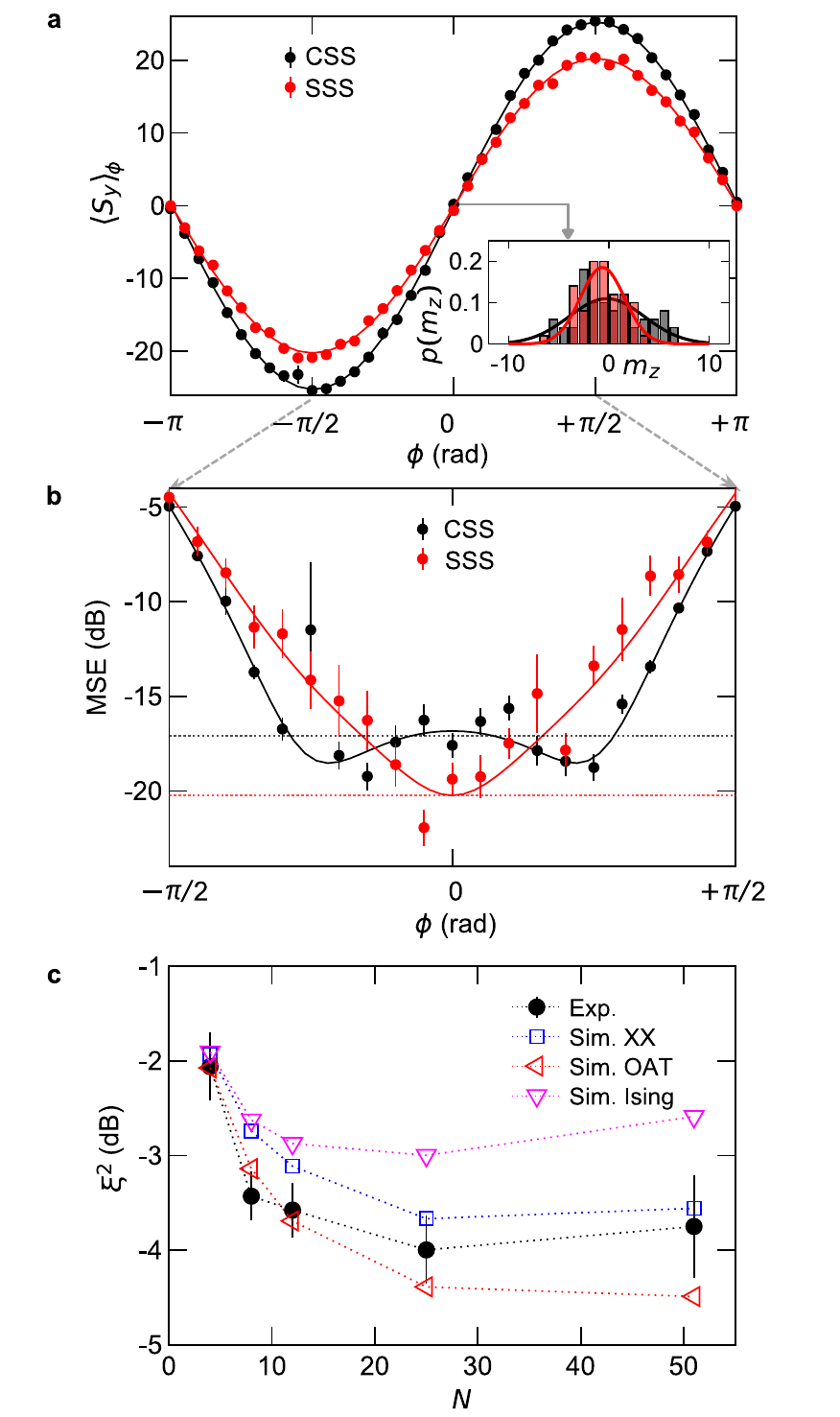} 
 \caption{\textbf{Phase estimation with a coherent spin state (CSS) and a spin-squeezed state (SSS) of 51 qubits.} a) Experimentally measured Ramsey fringe $\langle S_{y}\rangle_{\phi}$ (see methods \ref{met:measurements}) as a function of the phase $\phi$ imprinted onto the ions. (inset; probability to obtain the measurement outcomes $m_z$ after the Ramsey sequence for $\phi=0$). b) Mean squared error (MSE) Eq.~\eqref{eq:MSE} between imprinted and estimated phases as function of $\phi$. The dotted black line indicates the standard quantum limit (SQL) $1/N$ and the dotted red line indicates the gain in sensitivity of $~$3.2 dB for the SSS over the SQL. Solid lines are fits to the respective models (see methods~\ref{met:measurements}). c) Dependence of $\xi^2$ on the system size. The error bars are obtained via the Jack-knifing resampling method. We show comparisons to numerical calculations for the power-law XX model, as well as analytical calculations for the OAT model with $\chi = \overline{J}$ and for the power-law Ising model, all with effects of global dephasing taken into account (the dotted lines are a guide to the eye).}
 \label{fig:4}
\end{figure}

We plot theoretical calculations for the analogous OAT model (with $\chi = \overline{J}$) in the absence of decoherence for comparison, and observe that our measured phase space structure correlates well with the evolution of this fully collective model. Note that the approximate numerical methods we used before to determine the spin squeezing dynamics in the XX and TFI models in the presence of decoherence cannot be relied upon for such observables, leaving only our experimental results. Relative to the OAT model, the measured patches are smeared out in the horizontal direction, as a result of collective dephasing, but this smearing remains small enough to resolve the underlying separation of the Q-distribution. We also show the analogous results for the power-law Ising model in the absence of decoherence and with the same interaction range. Its quantum noise distribution beyond the Gaussian regime shows little resemblance in shape and magnitude to 
the OAT model (note very different scale in Fig.~\ref{fig:3}c) solely due to the finite interaction range and corresponding leakage of population out the permutationally-symmetric spin manifold at these late times.

While the Q-distribution features key signatures of the q-headed cat states, it does not indicate the presence of coherence between each of the observed coherent spin-like states. To explore this, we measure parity oscillations for the $q=2$ cat state with contrast $C$, which, in combination of the projection of our state onto the $\pm N/2$ eigenstates of $\hat{S}_z$ and the corresponding probabilities $p(m_z=\pm N/2)$ (see Fig.~\ref{fig:3}d,e), yields a measure of fidelity to the GHZ state $F=[p(m_z=N/2)+p(m_z=-N/2)+C]/2$. We observe the clear presence of parity oscillations in Fig.~\ref{fig:3}e. The generated state has a fidelity of $F=0.28\pm0.02$ to the GHZ state, which is below the required $0.5$ to certify $N$-partite entanglement. 
We attribute this relatively lack of fidelity primarily to decoherence in this system. Exact theoretical calculations in the absence of decoherence predict that fidelities of up to $0.916$ can be achieved for the 12-ion XX model with $\alpha\approx 1$
 (the fidelity drops to $0.865$ if the finite transverse field is taken into account). On the contrary, a decoherence-free Ising model with the same interaction range results in a fidelity of $<0.002$ at $T = \pi/q\overline{J}$. Thus, the use of spin exchange interactions offer significant room for GHZ state generation in non-collective systems --- a virtually impossible feat for finite-range Ising interactions.

\emph{Phase sensing with a 51-ion chain-} 
While we have demonstrated the production of SSS states with entanglement quantified by the Wineland parameter, we now study the performance of a SSS in a Ramsey interferometer and compare it to a CSS. For this study, we first prepare a SSS with our finite-range interaction and thereafter align the minimum variance axis of the variance ellipse perpendicular to the measurement axis ($z$-axis) by performing a rotation operation as described in methods \ref{met:measurements}. The phase $\phi$ is then imprinted on the spin vector by performing a rotation about the Bloch sphere $z$-axis. The projection of the spin along the $y$-axis shows sinusoidal variation of $\braket{\hat {S}_y}_{\phi}$, with $\hat {S}_y = \frac{1}{2}\sum_{j} \hat{\sigma}_j^{y}$, as a function of imprinted phases for both CSS and SSS cases, as shown in Fig.~\ref{fig:4}a. In the inset, we show the measured histogram of $m_z$, eigenvalues of the $\hat {S}_z$, from the single-shot outcomes in the measurement axis and demonstrate the narrowing of the distribution when a spin-squeezed state is utilized.
 
In practice, the measurement outcome of $m_z$ is used to estimate the imprinted phase $\phi$. We employ a linear estimator to find the imprinted phases on collective spin and calculate the mean-squared error (MSE) for the SSS and CSS. The results are presented in Fig.~\ref{fig:4}b. Notably, a reduction in the MSE is visible when the SSS is used over the CSS for small values of $\phi$ around $\phi =0$. For large values of $\phi$ the estimator poorly performs for both cases due to the fact our estimator is only unbiased for $\phi = 0$. 

Phase sensing can be improved by increasing the number of ions as the slope of the Ramsey fringes used in estimating the acquired phase increases with particle number as well as by reducing the variance via spin squeezing (see methods~\ref{met:measurements}). To this end, we show a scaling of the achievable spin squeezing as a function of the system size $N$, maintaining a fixed $J_0$; see Fig.~\ref{fig:4}c. The experimental results indicate an increase in $\xi^2$ for small $N$ but for large $N$ saturation occurs. In our system, the current limitation is imposed by slow entanglement generation and underlying dephasing channels, as we again confirm the optimal squeezing is similar to that achievable in an analogous OAT model with $\chi = \overline{J}$ and decoherence and dephasing taken into account. One could improve this by either increasing the spin-spin coupling strength or by reducing the dephasing effects. In fact, we already observe an improvement when the value of $J_0$ is doubled, while maintaining the same level of dephasing (see $N =12$, in Fig.~\ref{fig:1}c). Nonetheless, we also observe that compared to the corresponding power-law Ising model, our system still leads to a more robust generation of spin squeezing, owing to the relative stabilization of collective behavior as we have demonstrated.

\section{Conclusion}

The direct observation of the emergence of OAT collective behaviors in systems with finite-range interactions is a crucial step towards integrating entanglement into the best performing clocks operating with a large particle number. While we have demonstrated the utility of spin-exchange interactions to preserve spin alignment in a 1D chain, far better protection is expected to be achieved in optical qubits trapped in higher spatial dimensions~\cite{Perlin2020,bilitewski_dynamical_2021,Comparin2022,young_enhancing_2023,Block2023}, such as planar Coulomb crystals built via novel monolithic radiofrequency (rf) traps~\cite{Kiesenhofer2022,qiao_observing_2022},
or Penning traps~\cite{Bohnet2016,Itano1998}, as well as in optical tweezer arrays~\cite{Schine2022,Barredo2018} and 3D optical lattices~\cite{Campbell2017}. In these systems, it should be possible to work with arrays of a few hundred particles or more while enjoying single-particle control, under conditions where decoherence is limited and many-body interactions favoring spin alignment protect quantum states against external perturbations while generating scalable entanglement. In the case of trapped ions such conditions should be achievable with the aid of higher laser power. 
Furthermore, while squeezed states generated by OAT do not saturate the $1/N$ Heisenberg limit in standard Ramsey protocols, the corresponding interactions can be combined with time-reversal-based schemes~\cite{Davis2016,Colombo2022,Gilmore2021} to achieve a similar scaling. Moreover, the emergent collective dynamics of the XX model can be used to enrich the gate toolset for programmable quantum sensors~\cite{Marciniak2022, Kaubruegger2021}, and enhance their capability to measure time-varying frequencies close to the fundamental limit imposed by quantum mechanics in larger arrays.
Each of these generalizations of our current experiment or a combination of them could usher in a powerful new generation of entanglement-enhanced sensors. 

During completion of our work we became aware of related experiments using dressed Rydberg interactions in tweezer~\cite{Bornet2023,Eckner2023} and microtrap~\cite{Hines2023} array platforms.

\textbf{Acknowledgements.} We acknowledge stimulating discussions with members of the LASCEM collaboration about realizing spin squeezing in trapped ions with short-range interactions which initiated the project, as well as valuable feedback on the manuscript by W. Eckner and A. Kaufman. We also acknowledge support by the Austrian Science Fund through the SFB BeyondC (F7110) and funding by the Institut f\"ur Quanteninformation GmbH, by the Simons Collaboration on Ultra-Quantum Matter, which is a grant from the Simons Foundation, and by LASCEM via AFOSR No. 64896-PH-QC. Support is also acknowledged from the AFOSR grants FA9550-18-1-0319 and FA9550-19-1-0275, by the NSF JILA-PFC PHY-1734006, QLCI OMA-2016244, NSF grant PHY-1820885, by the U.S. Department of Energy, Office of Science, National Quantum Information Science Research Centers, Quantum Systems Accelerator, and by NIST.

\textbf{Author contributions}. The research was devised by SRM, MJ, RK, AMR and CFR. SRM and AMR developed the theoretical protocols. JF, MJ, FK, RB, and CFR contributed to the experimental setup. JF, MJ, and FK performed the experiments. MJ, JF and RK analyzed the data and SRM carried out numerical simulations. SRM, MJ, JF, RK, AMR, and CFR wrote the manuscript. All authors contributed to the discussion of the results and the manuscript.

\bibliography{IonSqueezing22}

\clearpage

\appendix
\renewcommand*{\thesubsection}{\Alph{subsection}}
\section*{Methods}

\label{sec:methods}
\subsection{Experimental platform}
\label{met:exp}
The experiment is performed on an analog quantum simulator based on trapped ions held in a macroscopic linear Paul trap~\cite{Kranzl2022b}. Two electronic states, namely S$_{1/2},m=+1/2$ and D$_{5/2},m=+5/2$, of a trapped $^{40}$Ca$^+$ ion form two pseudo-spin states $\ket{\downarrow}$ and $\ket{\uparrow}$, respectively. 

Long ion chains are held in the trap by a confining force in the radial directions, generated by a two-dimensional rf-quadrupole field that creates center-of-mass mode (COM) frequencies of about 2.93 MHz and 2.89 MHz. A set of dc voltages provides a weakly confining force along the direction of the rf-quadrupole and control the frequency splitting of the transverse COM modes. The axial trapping potentials are tuned in such a way as to achieve COM mode frequencies between $\omega_z =2 \pi \times 117$~kHz and $\omega_z =2 \pi \times 479$~kHz for ion numbers between 51 and 4. Before initializing the ions into the desired spin states, the ion chain's transverse motional modes are cooled close to the motional ground state via Doppler cooling and resolved sideband cooling techniques. Axial modes are sub-Doppler cooled via the polarization gradient cooling technique. 

A frequency-stable laser with a linewidth below 10~Hz and a wavelength of about 729~nm is used for coherent qubit manipulation by globally illuminating all ions from the radial direction with an elliptically shaped laser beam with a near spatial homogeneity. All ions are dissipatively initialized in the spin-down state. Thereafter, the collective spin vector is prepared along the $x$-axis by applying a global laser pulse that rotates the spin vector about the $-y$-axis. A composite pulse sequence is implemented to reduce the effects of the spatial inhomogeneity of the laser beam. More details can be found in the supplemental materials of ref.~\cite{Joshi2022}. 

\subsection{Generation of spin-spin interaction}
\label{met:interactions}
Spins are entangled via a two-tone laser field driving the spin and the radial motional degrees of freedom. The unwanted Stark shift generated from the coupling of the laser field to other electronic states is compensated by adding a third laser frequency component onto the two-tone field. This resultant field generates a global spin-spin interaction between the ions, representing a power-law decaying Ising-type interaction. In our experiments, the two-tone laser field is detuned by a frequency between 48 kHz and 25 kHz, for ion numbers between $4$ and $51$, from the highest frequency mode. The third laser beam is detuned by $+1.4$ MHz from the carrier transition. The center frequency of the two-tone field is detuned from the spin resonance to engineer a transverse field Ising interaction, 
\begin{align}\label{eq:XplusZ}
\hat{H}_{\rm{PL-TFI}} = \sum_{i<j} \frac{J_0}{|i-j|^\alpha} \hat{\sigma}_i^x \hat{\sigma}_j^x + B \sum_{i} \hat{\sigma}_i^z.
\end{align}
The transverse component strength is $B = 9500$ rad/s while $J_0\le 560$ rad/s, thus allowing us to transform the above interaction into the XX interaction in the frame rotating with the transverse field, by neglecting the resulting fast oscillating terms (rotating wave approximation).

\subsection{Mitigation of global dephasing effects}
\label{met:echo}
The fluctuation of laser phase and ambient magnetic fields incur dephasing of the spins during the preparation of the spin-squeezed state. The magnetic field fluctuations are predominately caused by the current flowing in the electrical appliances in the laboratory thus having contribution at 50~Hz and its higher-order harmonics. Further details of mitigating these magnetic field fluctuations via a feed-forward method are presented in ~\cite{Kranzl2022b}.
The remaining dephasing induced by the slow phase variation of the laser field and the transition frequency change is reduced in our case by the spin-echo technique and we observe an improvement in $T_2$ coherence time from $42\pm 2$~ms to $68\pm 6$~ms after implementing the spin-echo scheme. 

\subsection{Measurements}
\label{met:measurements}
\begin{figure*}[ht]
\renewcommand{\thefigure}{ A1}
\centering
 \includegraphics{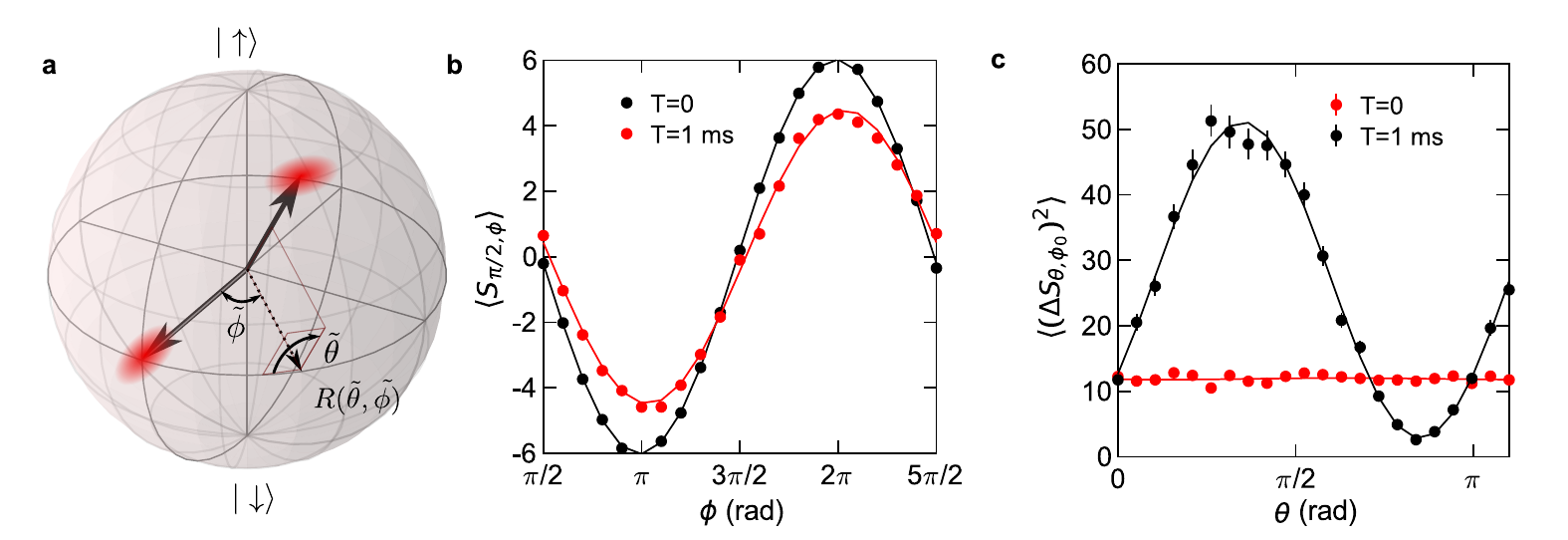}
 \caption{\textbf{Assessment of the experimentally prepared spin state} a) The outcome of the entangling interaction on the spins pointing along the $x$-axis is depicted as a variance ellipse. $\xi^2$ is evaluated from two sets of measurements that are obtained after a rotation $R(\tilde{\theta},\tilde{\phi})$ has been applied to the state. b) Applying $R(\pi/2,\tilde{\phi})$ for various values of $\tilde{\phi}$ allows us to measure the spin projection $\langle S_{\pi/2, \phi}\rangle$ in any direction along the equator. From the sinusoidal fits (solid lines) we obtain the Bloch vector length and orientation and the angle $\phi_0$ between $x$-axis and the mean spin orientation. c) Applying $R(\tilde{\theta}, \phi_0)$ for various values of $\theta$ allows us to measure the variance $\langle (\Delta S_{\theta, \phi_0})^2 \rangle$ in any direction orthogonal to the mean spin direction. From the fitted data we extract the minimal orthogonal variance $\min_{\mathbf{n}_{\perp}}\braket{(\Delta\hat{S}_{\mathbf{n}_{\perp}})^2} $ and the angle $\theta_0$ for which minimal variance is aligned with the $z$-axis. }
 \label{fig:S1}
\end{figure*}

In order to evaluate the various quantities presented in this manuscript we perform a collective rotation of the spins
\begin{align}
 R(\tilde{\theta},\tilde{\phi}) = \prod_j e^{-i\frac{\tilde{\theta}}{2} (\sigma_j^x \cos\tilde{\phi}+\sigma_j^y \sin\tilde{\phi})},
\end{align}
as sketched in   Fig.~\ref{fig:S1}a. The spins are rotated by an angle $\tilde{\theta}$ around an arbitrary axis in the $xy$-plane that is parameterized by $\tilde{\phi}$. Followed by a projective measurement in the $z$-basis this allows us to measure the collective spin operators in any basis $\hat{S}_{\theta, \phi} = R^{\dagger}(\tilde{\theta},\tilde{\phi})\,\hat{S}_z\,R(\tilde{\theta},\tilde{\phi})$ characterized by polar/azimuthal angles $\theta/\phi$ on the collective Bloch sphere. Here, the polar/azimuthal angles are related to rotation angles by,  $\phi = \tilde{\phi}+\pi/2$ and $\theta=\tilde{\theta}$.

\emph{Spin squeezing parameter --} In order to estimate the spin squeezing parameter Eq.~\eqref{eq:Sqparam} we perform a set of measurements where either $\tilde{\phi}$ or $\tilde{\theta}$ are scanned. The evolution under the TFI model changes the orientation of the collective spin vector in contrast to the XX model thus consecutive laser pulses are needed to be operated under the knowledge of this phase accumulation. In our experiments, the transverse field is engineered by detuning the laser field by the strength $B \gg J_0$ such that the RWA is met, while the spin vector rotation is accounted for by the phase evolution of the detuned laser field during the operation of TFI interaction. 
Nonetheless, we still observe small changes in the spin orientation within the $xy$-plane, which can arise due to changes in the transition frequency from unaccounted factors. In order to characterize the length of the spin vector and its orientation we perform a set of measurements $\langle \hat{S}_{\pi/ 2, \phi}\rangle $ where the measurement basis is changed within the $xy$-plane.
These quantities are estimated by fitting the experimental data (see   Fig.~\ref{fig:S1}b for two representative fits) by a $sin$ function. The length of the spin vector is given by the contrast of the fit and phase offset ($\phi_0$) represents the angle between the spin vector and the $x$-axis.

In a subsequent series of measurements, we measure $\langle \big(\Delta \hat{S}_{\theta, \phi_0}\big)^2\rangle$ for different $\theta$, which corresponds to a scan of the variance in the plane orthogonal to the mean spin direction. 
In order to determine the minimal orthogonal variance and the angle at which this variance is aligned with the $z$-axis we fit $\langle \big(\Delta \hat{S}_{\theta, \phi_0}\big)^2\rangle$ by a fit function of the form 
$V(\theta) = (V_{\rm max} - V_{\rm min}) \sin^2(\theta - \theta_0) + V_{\rm min}$ (see   Fig.~\ref{fig:S1}c for two representative fits), so that $V_{\rm min} = 4\min_{\mathbf{n}_{\perp}}\langle (\Delta\hat{S}_{\mathbf{n}_{\perp}})^2\rangle $ 

The same sets of measurements can be used to evaluate the correlation functions in the three Cartesian bases underlying the results presented in Fig.~\ref{fig:1}d, and Fig.~\ref{fig:2}. Correlations in the $x(y)$-basis are evaluated from the data set with $\tilde{\theta} = \pi / 2$ for $\tilde{\phi} = 3\pi / 2 (0)$ and correlations in the $z$-basis are evaluated from the second set with $\tilde {\phi}= \phi_0$ at $\theta = 0$.

\emph{Husimi Q-distribution --} The Husimi $Q(\theta, \phi)$-distributions visualized in Fig.~\ref{fig:3} are obtained by measuring the overlap between the state $\ket{\psi}$ after the final rotation $R(\tilde{\theta},\tilde{\phi}) $ and the maximally polarized state $\ket{\downarrow,\dots, \downarrow}$, so that 
\begin{align}
 Q(\theta, \phi) = |\bra{\downarrow,\dots, \downarrow}R(\tilde{\theta},\tilde{\phi})  \ket{\psi}|^2,
\end{align}
which is equivalent to measuring the overlap with a coherent spin state $\ket{\bf{n}(\theta, \phi)} = R^{\dagger}(\tilde{\theta},\tilde{\phi})  \ket{\downarrow,\dots, \downarrow}$. 

\emph{Characterization of the 2-headed cat state -- }
In order to characterize the 2-headed cat state in Fig.~\ref{fig:3}d, e we first rotate the cat state with a pulse $R(\pi /2, \phi_1)$ where $\phi_1$ is chosen such that the axis of the cat state is aligned with the $z$-axis, i.e. the Husimi distribution of the aligned state is maximal at the north and south pole. The probability distribution $p(m_z) = |\braket{m_z|\rm cat}|^2$ in Fig.~\ref{fig:3}d is obtained by projecting the aligned cat state onto the magnetization eigenstates $\hat{S}_z\ket{m_z} = m_z \ket{m_z}$. 
In order to evaluate the parity oscillation ins Fig.~\ref{fig:3}e we perform another rotation pulse $R(\pi/2,\tilde{\phi})$ and measure the parity according to $\langle \hat{P}_{\phi}\rangle = \sum_{m_z} e^{-i \pi (S+m_z)}\,|\langle m_z | R(\pi/2,\tilde{ \phi}) \ket{\rm cat}|^2$. The phase factor $e^{-i \pi (S+m_z)}$ is $+1(-1)$ if the number of ions in the excited is even(odd).  

\emph{Phase sensing experiment --} 
In Fig.~\ref{fig:4} we study the metrological utility of the SSS we have prepared in terms of a Ramsey experiment. In order to perform a Ramsey experiment the minimum variance direction of the SSS has to be aligned with the $y$-axis before the phase $\phi$ to be sensed is imprinted according to a unitary $U(\phi) = e^{-i \phi \hat{S}_z}$ before the last step a $\pi/2$ pulse around the $x$-axis is applied and the ions are projectively measured. We can combine phase imprinting $R_z = e^{-iS_z \phi}$ and the final measurement pulse $R(\pi/2,0)$ in a single rotation such that the Ramsey fringes are given by $\langle S_{y}\rangle_{\phi} = - \langle R^{\dagger}(\pi/2,\phi) S_z R(\pi/2,\phi)\rangle$. 

In order to study the sensitivity of our spin-squeezed sensor we consider an estimator $\phi_{\rm est}(m) = m / |\bm{\hat{S}}|$ that estimates the phase based on a single measurement of the magnetization $m$. We extract $|\langle\bm{S}\rangle|$ from the Ramsey fringe. For this estimator the mean squared error (MSE) can be expressed in terms of measurable expectation values according to 
\begin{align}\label{eq:MSE}
 {\rm MSE}(\phi) &= \sum_{m = - N / 2}^{N / 2} \left(\phi - \phi_{\rm est}(m)\right)^2 p(m|\phi)\notag \\
 & = \phi^2 - \frac{2 \phi \langle S_{y}\rangle_{\phi}}{|\langle \bm{S}\rangle |} + \frac{\langle (S_{y})^2 \rangle_{\phi}}{|\langle \bm{S}\rangle |^2}
\end{align}
Here the conditional probability is given by $p(m|\phi) = |\bra{m} S_y \ket{\psi_{\phi}}|^2$. The experimental data in Fig. \ref{fig:4}b are fitted to a function MSE$(\phi)=(\phi^2+ a_1 \phi \sin\phi+ a_2 \sin^2\phi+ a_3 \cos^2\phi )$ after generalising Eq.~\eqref{eq:MSE}. We use the fit parameter $a_3$ to determine the smallest MSE at $\phi = 0 $ to be  $10\log_{10} (a_3)$ for a CSS and SSS respectively.

\subsection{Error estimation in the measurements}
\label{met:error}
In the present works, each measurement point is an average of 50 to 600 repetitions of experimental realizations.
Measurement error bars are produced everywhere in this manuscript using the Jackknife resampling method~\cite{Sen1968}, except for estimating the total spin vector length $\braket{\hat{S}^2}$ in Fig.~\ref{fig:1}d. Here, instead, the error propagation formula is used to estimate the error based on the standard errors of the single components.

\subsection{Numerical methods}
\label{met:numerics}
For calculations involving evolution under the OAT model, we can exactly solve the wave equation for virtually any $N$, taking advantage of the permutation
symmetry in this model. For the power-law Ising model, we can likewise solve for the full, exact dynamics for small $N$, while exact analytical results are available for the dynamics of arbitrary one- and two-body observables~\cite{foss-feig_nonequilibrium_2013}.

To include the effects of a white-noise global dephasing process in the dynamics, we utilize the master equation formalism to model the dynamics of density operator $\hat{\rho}(T)$ of the state as
\begin{align}
\begin{split}
 \frac{\partial\hat{\rho}(T)}{\partial T}& = -i\left[\hat{H},\hat{\rho}(T)\right]\\
 &+ \Gamma_z\left(\hat{S}_z\hat{\rho}(T)\hat{S}_z - \frac{1}{2}\left\{\hat{S}_z^2,\hat{\rho}(T)\right\}\right)
 \end{split}
\end{align}
for any of the Hamiltonians $\hat{H}$ we model in the main text, where we assume the global dephasing rate is given by $\Gamma_z = 2/T_2$. To solve for the corresponding dynamics, we note that the jump operator $\hat{S}_z$ commutes with the OAT Hamiltonian, as well as the power-law Ising and XX Hamiltonians, and thus we can directly compute its effects on the resulting observables and correlators after solving for the unitary dynamics. This can be performed exactly by the following replacements:
\begin{align}
 \braket{\hat{S}_{x/y}(T)} &\rightarrow e^{-\Gamma_z T/2}\braket{\hat{S}_{x/y}(T)}\\
 \braket{\hat{S}_{x/y}\hat{S}_z(T)} &\rightarrow e^{-\Gamma_z T/2}\braket{\hat{S}_{x/y}\hat{S}_z(T)}\\
 \braket{(\hat{S}_{x}^2 - \hat{S}_{y}^2)(T)} &\rightarrow e^{-2\Gamma_z T}\braket{(\hat{S}_{x}^2 - \hat{S}_{y}^2)(T)}\\
 \braket{\{\hat{S}_{x},\hat{S}_{y}\}(T)} &\rightarrow e^{-2\Gamma_z T}\braket{\{\hat{S}_{x},\hat{S}_{y}\}(T)},
\end{align}
for the anticommutator $\{\cdot,\cdot\}$, where $\Gamma_z$ is the rate of global dephasing.

For modeling the dynamics of the transverse-field Ising model in Eq.~\eqref{eq:XplusZ}, as well as the corresponding XX model in the large transverse-field limit, we must resort to more efficient, approximate schemes for larger $N$. Furthermore, we must explicitly solve for the effects of the global dephasing in the dynamics for the transverse-field Ising model, as the corresponding jump operator no longer commutes with the Hamiltonian dynamics. To do this, we utilize the dissipative discrete truncated Wigner approximation (DDTWA)~\cite{Schachenmayer2015,Zhu2019,huber_realistic_2022} to efficiently solve for the quantum dynamics when $N>10$ for these two models (otherwise, we resort to exact methods). DDTWA has previously been benchmarked for calculations of quantum spin dynamics and spin squeezing generation for various models~\cite{young_enhancing_2023,barberena_fast_2022}, and affords an efficient semiclassical description of the dynamics.

To do this, we introduce classical variables $\mathcal{S}_i^\mu$ corresponding to the value of $\braket{\hat{\sigma}_i^\mu}$. For an initial spin-polarized state along $+x$, for example, we form a discrete probability distribution (Wigner function)
\begin{align}
\begin{split}
 W(\bm{\mathcal{S}}_i) =& \frac{1}{4}\delta(\mathcal{S}_i^x - 1)\Big[\delta(\mathcal{S}_i^y - 1) + \delta(\mathcal{S}_i^y + 1)\Big]\\
 &\times\Big[\delta(\mathcal{S}_i^z - 1) + \delta(\mathcal{S}_i^z + 1)\Big].
 \end{split}
\end{align}
This corresponds to the four phase space points $(\mathcal{S}_i^x,\mathcal{S}_i^y,\mathcal{S}_i^z) = (1,1,1)$, $(1,1,-1)$, $(1,-1,1)$, $(1,-1,-1)$, each occurring with equal probability $1/4$ for each spin. We can then solve the coherent dynamics by evolving these variables via the corresponding mean field equations for the relevant Hamiltonian, combined with randomly sampling initial values for $(\mathcal{S}_i^x,\mathcal{S}_i^y,\mathcal{S}_i^z)_{1\leq i\leq N}$ according to the above distribution, independently for each $i$. For an ensemble of such trajectories, quantum expectation values may then be approximated by $\langle \hat{\sigma}_i^\mu(t)\rangle \approx \overline{\mathcal{S}_i^\mu(t)}$, where $\overline{\,\cdot\,}$ denotes averaging over all trajectories. We can also compute symmetrically-ordered correlators via $\langle(\hat{\sigma}_i^\mu\hat{\sigma}_j^\nu + \hat{\sigma}_j^\nu\hat{\sigma}_i^\mu)(t)\rangle/2 \approx \overline{\mathcal{S}_i^\mu(t)\mathcal{S}_j^\nu(t)}$. This averaging can lead to effects beyond mean-field theory arising from the underlying quantum noise distribution, owing to the generic nonlinear nature of the mean-field equations for an interacting system.

To model the effects of the global dephasing, we can include a stochastic contribution to our mean field equations, see~\cite{huber_realistic_2022,young_enhancing_2023}. For example, for evolution under the XX interaction in Eq.~\eqref{eq:XYInteraction}, the resulting equations of motion for our classical variables are then given by the set of Stratonovich stochastic differential equations
\begin{align}
d\mathcal{S}_i^x &= -\sum_{j\neq i} J_{i,j}\mathcal{S}_i^z\mathcal{S}_j^y dT - \sqrt{\Gamma_z}\mathcal{S}_i^ydW^z\\
d\mathcal{S}_i^y &= \sum_{j\neq i} J_{i,j} \mathcal{S}_i^z\mathcal{S}_j^x dT + \sqrt{\Gamma_z}\mathcal{S}_i^xdW^z\\
d\mathcal{S}_i^z &= \sum_{j\neq i} J_{i,j} \left(\mathcal{S}_i^x\mathcal{S}_j^y - \mathcal{S}_i^y\mathcal{S}_j^x\right) dT.
\end{align}
Here, $dW^z \equiv dW^z(T)$ is a Wiener increment such that $\langle dW^z(T)dW^z(T)\rangle = dT$, and $\langle dW^z(T)\rangle = 0$~\cite{Gardiner2009}. To properly model the experiment, we directly utilize coupling matrices $J_{i,j}$ as characterized in our platform for all calculations, see e.g. Fig.~\ref{fig:1}b. We average our results over $5 \times 10^3$ trajectories, which we find to be sufficient to obtain a sampling error well within the size of experimental error bars.

\subsection{Linear spin wave}
\label{met:LSW}
In Fig.~\ref{fig:2} we study the generation SWE by the use of the Holstein-Primakoff approximation, which maps spin operators into quadratures written in terms of bosonic annihilation and creation operators $\hat{b}_i$ and $\hat{b}_i^\dagger$. In the linear spin wave approximation we have
\begin{align}
\hat{\sigma}_i^x & = 1 - 2 \hat{b}^{\dagger}_i \hat{b}_i \\
\hat{\sigma}_i^y & \simeq - i (\hat{b}_i - \hat{b}_i^{\dagger}) \\
\hat{\sigma}_i^z & \simeq - (\hat{b}_i + \hat{b}_i^{\dagger}). 
\end{align}
The linear approximation is valid if $\langle \hat{\sigma}_i^x\rangle \lesssim 1$, i.e. the spins remain primarily polarized along the $x$-axis. The initial CSS polarized along $+x$ maps to the bosonic vacuum state, and the dynamical evolution under Eq.~\eqref{eq:XYInteraction} describes the generation of SWE in terms of these bosonic fields.
 
In momentum space the bosonic operators are given by $\hat{b}_{k} = \sum_{i} e^{- i k r_i }\hat{b}_{i}/\sqrt{N}$, where $r_i$ is the location of ion $i$ in the chain. We wish to study the occupation $\braket{ \hat{n}_{k}} = \braket{ \hat{b}_{k}^{\dagger}\hat{b}_{k} }$ of the different modes with momentum $k$. The mode occupation can be expressed in terms of the spin correlations that are measured in the experiment
\begin{align}
\frac{\braket{ \hat{n}_{+k} } + \braket{ \hat{n}_{-k} }}{2} \simeq &\frac{1}{2N} \sum_{i<j} \braket{\hat{\sigma}_i^y\hat{\sigma}_j^y + \hat{\sigma}_i^z\hat{\sigma}_j^z }\cos(k (r_i - r_j))\notag \\ & - \frac{1}{2N} \sum_i \braket{ \hat{\sigma}_i^x} + \frac{1}{2} = \langle \hat{C}_k \rangle.
\end{align}

\end{document}